\documentstyle[aps,psfig]{revtex}

\draft 

\def\rem #1{{}}

\def\dd{\displaystyle} 
\newcounter{bidon}
\def\FPA{{\em FPA }}
\def\TBS{{\em TBS }}
\def\TSS{{\em TSS }}
\def\NTIII{{\em NT30}}
\def\NTVI{{\em NT60}}

\def\ie{{\em i.e. }}

\author{ G. ORON and H. J. HERRMANN } 
\title{Exact calculation of force networks in granular piles} 
\address{Laboratoire de Physique et M\'ecanique des Milieux H\'et\'erog\`enes, \'Ecole de
  Physique et de Chimie Industrielles de la ville de Paris, 10 rue
  Vauquelin, 75231 Paris Cedex 05, FRANCE \\
  email : oron@pmmh.espci.fr}

\begin{document}
\maketitle
\begin{abstract}
  We present calculations of forces for two dimensional static
  sandpile models.  Using a symbolic calculation software we obtain
  exact results for several different orientations of the lattice and
  for different types of supporting surfaces.  The model is simple,
  supposing spherical, identical, rigid particles on a regular
  triangular lattice, without friction and with unilateral spring-like
  contacts.  Special attention is given to the stress tensor and
  pressure on the base of the pile.  We show that orientation of the
  lattice and the characteristics of the supporting surface have a
  strong influence on the physical properties of the pile.  Our
  results agree well with numerical simulations done on similar
  systems and show, in some specific cases, a dip \ie a depression
  under the apex of the pile.  We also estimate that the algorithm we
  have developed can be easily adapted to other configurations and
  models of granulates and can be used in other physical cases where
  piecewise linear systems are encountered.
\end{abstract}

  



\section{Introduction}
Extensive interest has been devoted to the study of granular matter in
the last few years~\cite{jaeger96}. It exhibits many surprising
phenomena and even though classical mechanics is a mature domain,
granulates in general, and static granular systems in particular still
pose many open questions.  In the case of a static heap of grains it
was observed that the pressure on the supporting surface presents a
local minimum (a {\em dip}) under the apex of the pile rather than the
intuitively expected maximum~\cite{smid81,huntley97}.  This
observation is qualitatively explained by an arching effect,
transporting the charge of the of the grains' weight to the sides of
the pile~\cite{edwards96b}. This idea was later incorporated in a new
continuum
approach~\cite{bouchaud95,bouchaud95b,wittmer97,cargese-bouchaud}
under the so called ``fixed principal axis'' hypothesis model ({\em
  FPA}), in which the constitutive equation needed to close the system
of equations states that the the stress tensor has its principal axis
always pointing in the same direction. It is claimed that this
characteristic is ``remembered'' by the grains from the moment they
were buried at the surface of the pile when the pile has grown.  This
model gives results in very good agreement with the experimental data
of 3D piles.  The validity of this approach is not generally accepted,
and some authors~\cite{savage97,cargese-savage} reject the basic
assumption of the \FPA model, pointing out that experiments on wedge
form piles does not show any dip as predicted by the \FPA model and
that the \FPA model cannot account for the observed sensitivity of the
force networks in the heap to the base boundary conditions (see
Ref.~\cite{savage97,cargese-savage} and references herein).  It is
claimed that the well known elasto-plastic continuum model of soils is
still valid in the pile's case. The dip, in this model is accounted by
the existence of regions with different constitutive behaviors,
plastic in the outer region and elastic in the inner
region~\cite{cantelaube97,cargese-cantelaube}.

Even though continuum models are very useful they do not give
many clues about the micro-mechanical origin of the granular behavior,
since these models are homogenized models taking a mean over a
large number of grains.  Some micromechanic models were proposed in
order to link the grain-size physics to the pile-size phenomena~%
\cite{stauffer87,liffman92,hong93,claudin97,huntley93,liffman94,hemmingsson95,hemmingsson97,luding97}.
One simple model is an array of rigid spheres arranged on a diamond
lattice~\cite{hong93}. Under this kind of pile the normal force was
shown to be constant even when periodic vacancies are introduced
throughout the pile~\cite{huntley93}. Varying the sizes of the
grains or introducing attractive forces might change the forces under
the pile, but still does not reproduce the experimental force
profile~\cite{liffman94}.

Many of the discrete models used for granular matter use a large
variety of numerical simulation techniques like molecular
dynamics~\cite{luding97,liffman92}, contact dynamics~\cite{radjai96b},
cellular automata~\cite{hemmingsson95,hemmingsson97} and others.
Although these algorithms are useful in reproducing many of the
characteristics of granular matter, none of these methods is
completely satisfactory. Convergence is very slow due to the highly
nonlinear character of the contact forces. The amplitude of forces in
granular matter spawns several orders of
magnitude~\cite{liu95,jaeger96} producing badly conditioned numerical
systems, and granular media are extremely sensitive to
fluctuations~\cite{claudin97,clement97}. As a result, much care should
be taken when numerically simulating granulates since cumulative
roundoff errors might give rise to errors of considerable amplitude.
Moreover, besides some very simple
models~\cite{hong93,huntley93,menon}, analytical results are uncommon
for these discrete models so that numerical results are seldom
verified.

A symbolic calculation software allows, in some cases, to perform
numerical-like calculations while avoiding any numerical errors. When
using such software one can perform automated analytical calculations
as if they were done by hand but on systems of a size too large to be
calculated by a human.

In this paper we present an implementation of such a calculation, in
the case of a pyramidal piling of 2D discs, under the effect of
gravity, and in the absence of friction. We use ``spring-like''
contacts under compression between discs and suppose them having
infinite rigidity, solving the equilibrium equations in a
straightforward way. We do not need any damping at the contacts, as it
is often the case in numerical methods, since no dynamics is used in
order to find the solution. Results obtained with this scheme are free
of any numerical error.

We study a wide variety of geometrical configurations, with different
supporting surfaces, lattice orientations and external constraints. In
particular, some of the configurations proposed in Ref.~\cite{luding97},
and studied using molecular dynamics, will be treated so to provide a
cross check of numerical results.

The generality of this method makes it versatile and easily adaptable
to other static cases like silo geometries, 3D pilings of spheres or
the study of the force network under external mechanical constraints
or any other case where piecewise linear equation systems are to be
resolved.

The outline of the paper is the following: The model is exposed in
Sec.~\ref{sec:model} and the different configurations are listed in
Sec.~\ref{sec:configs}. The algorithm is detailed in
Sec.~\ref{sec:alg}.  Sec.~\ref{sec:results} present the results for
the different realizations that are discussed in
Sec.~\ref{sec:conclusions}.

\section{The model}
\label{sec:model}
Since we aim at getting exact results we must consider a simple model.
First, we will consider only two dimensional piles formed of
identical, frictionless, spherical grains (discs), of same radius $R$
and weight $w$.  Those grains will be always arranged on a horizontal
surface \rem{!! Rajoute horizontal !!} to form a pile on a triangular
lattice, which is the natural lattice in this case.  In this case the
coordination number $n_c$ (the number of neighbors or the number of
possible contacts) is greater than twice the space dimension so that
if one only considers infinitely stiff discs, the system is
hyperstatic \ie the number of degrees of freedom in the system is
bigger than the number of equations imposed by the mechanical
equilibrium and the stability of the system $ ( \sum \vec F = 0 ) $ is
not sufficient to determine all forces. In such case, one must
introduce new relations in order to close the problem.

The easiest way to introduce the necessary relations is to consider
elastic, spring-like contacts between the discs, replacing the pile by
an array of point-like masses linked by springs, so that the force
applied by disc $ d_1 $ on disc $ d_2 $ is:

\begin{equation} 
\label{eq:Hook}
\vec F_{d_1 \rightarrow d_2} =-\frac{1}{\tau} (2 R-|\vec r_{d_1}- 
\vec r_{d_2}|)
\, \frac{\vec r_{d_1}-\vec r_{d_2}} {|\vec
  r_{d_1}-\vec r_{d_2}|} \theta(2 R-|\vec r_{d_1}-\vec r_{d_2}|) 
\label{eq:force}
\end{equation}

\noindent where $ \tau $ is the inverse of the elastic modulus (or the
softness) of each disc-disc contact, and $\vec r_{d_n}$ the
position of the disc $d_n$.  $ \theta $ represents the Heaviside step
function.  The $\theta $ functions are introduced in order to take
into account the unilateral character of the contact forces in dry
granular matter, \ie that the discs can push (when $2 R-|\vec
r_{d_1}-\vec r_{d_2}|>0$) but not pull each other.  The system is
therefore non linear and clearly impossible to solve analytically,
with or without the use of a symbolic calculation software, so some
kind of simplifying assumption must be introduced. In particular, we
consider the case of hard discs;
\begin{equation}
 \tau \ll R/w \label{eq:tausmall} 
\end{equation}
Since forces are finite, this condition implies that the displacements
of the discs from the lattice points must tend to $0$ with $ \tau $ , hence, when calculating
to first order in $ \tau $, these displacements are proportional to
$\tau$. We can develop the equations to first order in all the
displacements, linearizing the $(2 R-|\vec r_{d_1}-\vec r_{d_2}|) \,
\frac{\vec r_{d_1}-\vec r_{d_2}} {|\vec r_{d_1}-\vec r_{d_2}|}$ terms.
The remaining non-linear parts involved are the $ \theta $ functions
that cannot be linearized around 0.

Dealing with the non-linearity arising from the $\theta $ function is
the hardest part of the solution. One may consider calculating all
possible combinations of absent/present contacts, solve the linear
system obtained for each one of them and keep only solutions
consistent with the conditions of the originally supposed contact
network.  Unfortunately, this method is extremely time consuming,
since one must solve for $ 2^c $ different realizations, with $c$ the number of
contacts in the system. We will deal with this difficulty by an
iterative method described in Sec.~\ref{sec:alg}.

Although we can use this algorithm for any value of $\tau$ so that
$\tau\ll R/w$ (slightly deformable grains), in the following we will
restrain ourselves to the case of rigid discs, taking (analytically)
the limit $\tau\rightarrow 0$, in order to get results independent of
$\tau$.

A question we did not address in this work, is the existence and
uniqueness of the solution. Since the system of equations is
non-linear both questions are open and deserve further investigation.
We always found a solution for whatever pile size we studied, so that
the existence is assured, at least for the pile sizes we have
investigated.

\section{The geometrical configurations}
\label{sec:configs}
\begin{figure}[htb]
  \centerline{
    \psfig{figure=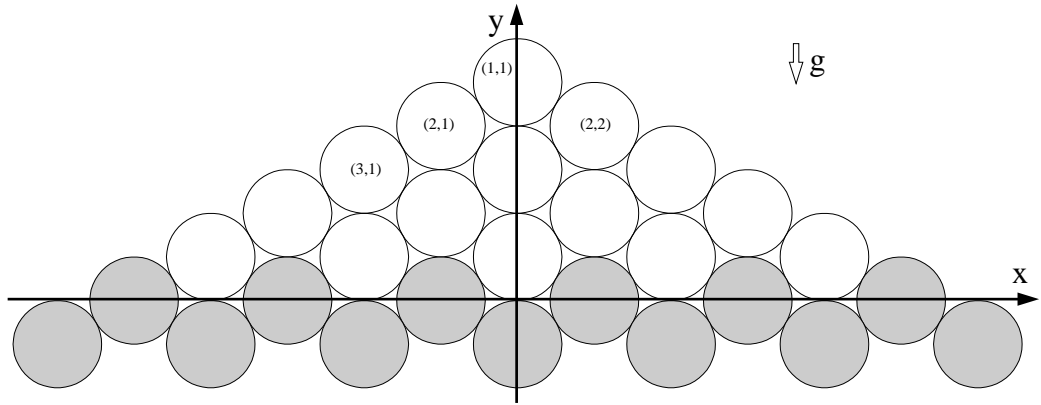,width=14cm} }
  \caption[A ``tilted'' triangular lattice pile of 5
  layers ({\em TBS case})]{A ``tilted'' triangular lattice pile of 5
    layers ({\em TBS case}) with a slope of $30^\circ$, the 2 gray
    layers represent the ``bumpy'' floor. Those two layers have their
    centers fixed on the lattice points. The indexing convention
    $(i,j)$ used is shown.  }
  \label{fig:TBS-pile}
\end{figure}

\begin{figure}[htb]
  \centerline{
    \psfig{figure=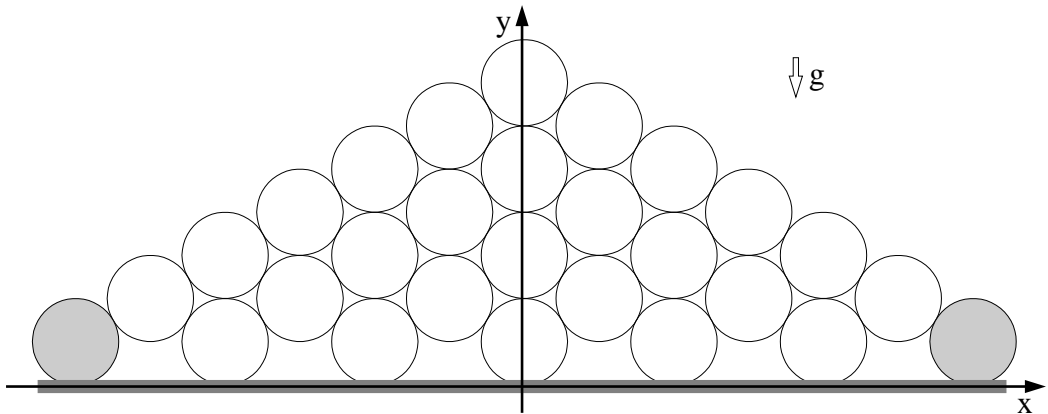,width=14cm} }
  \caption[A ``tilted''
  triangular lattice pile of 7 layers ({\em TSS case})]{A ``tilted''
    triangular lattice pile of 7 layers ({\em TSS case}) with a slope
    of $30^\circ$, the 2 gray discs (corner stones) have their centers
    fixed on the lattice points.}
  \label{fig:TSS-pile}
\end{figure}

\begin{figure}[htb]
  \centerline{ \psfig{figure=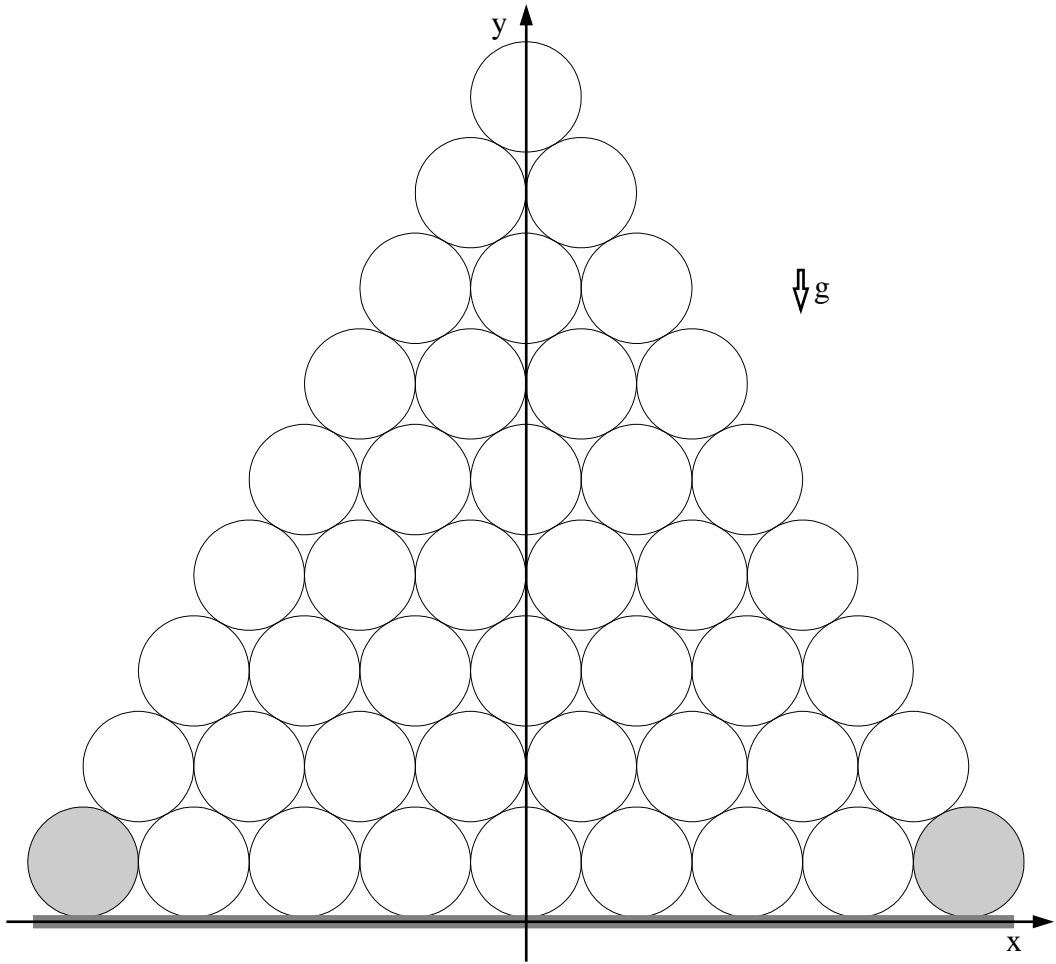,width=14cm} }
  \caption[A triangular lattice pile of 9 layers in the {\em NT60}
  case]{A triangular lattice pile of 9 layers in the {\em NT60} case,
    the 2 gray discs (corner stones) are fixed on the lattice points.}
  \label{fig:NT60-pile}
\end{figure}

\begin{figure}[htb]
  \centerline{ \psfig{figure=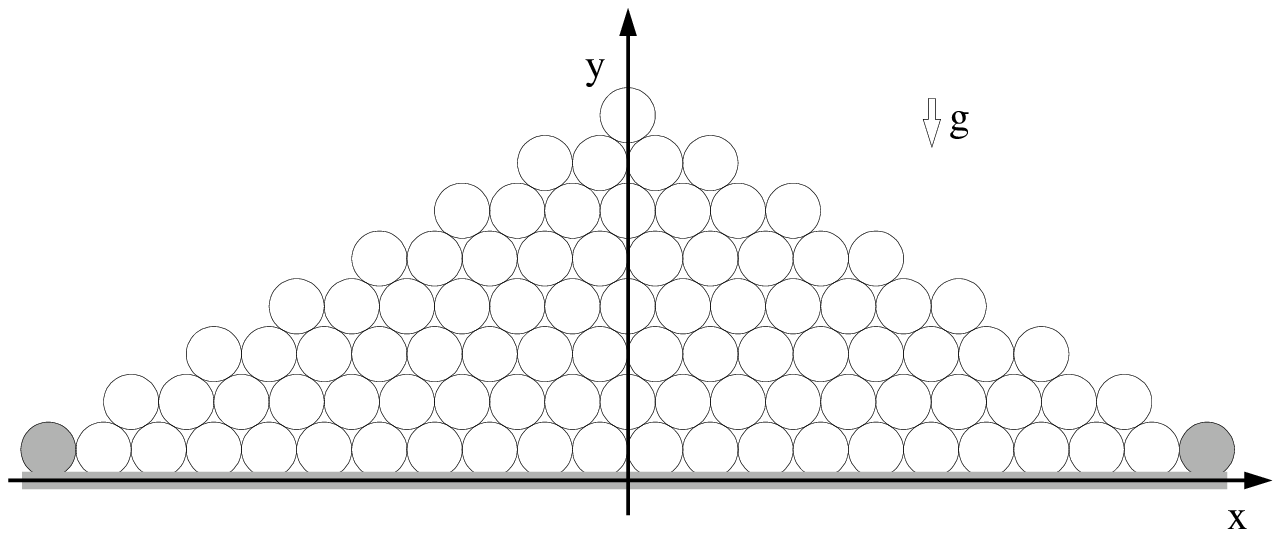,width=14cm} }
  \caption[A triangular lattice pile of 8 layers with $30^\circ $
  slope (the {\em NT30} case)]{A triangular lattice pile of 8 layers
    with $30^\circ $ slope (the {\em NT30} case), the 2 gray discs
    (corners stones) are fixed on the lattice points.}
  \label{fig:NT30-pile}
\end{figure}

We have studied the following geometrical configurations:

\begin{itemize}
\item A ``tilted''\footnote{In the following we will use ``tilted'' to
    designate a lattice of the type shown in Fig.~\ref{fig:TBS-pile}
    and~\ref{fig:TSS-pile}. The term ``untilted'' will refer to a
    lattice of the type shown in Fig.~\ref{fig:NT60-pile}
    and~\ref{fig:NT30-pile}.}  triangular lattice pile with $30^\circ$
  slope and the following surface conditions:

  \begin{itemize}
  \item ``bumpy'' surface, as shown in Fig.~\ref{fig:TBS-pile}. The
    pile poses on top of two layers of fixed discs; the centers of
    those discs are maintained on the triangular lattice points in
    order to simulate an infinitely rough surface. \rem{!!!!} We will
    refer to this case as the {\em TBS} case ({\em TBS} for {\em
      T}ilted {\em B}umpy {\em S}urface).
    
  \item ``smooth'' flat surface with only the outermost base discs
    fixed, as shown in Fig.~\ref{fig:TSS-pile}. In the
    following will refer to these discs as the {\em corner stones}.
        This case will be refered to as the {\em TSS} case. ({\em TSS} for
    {\em T}ilted {\em S}mooth {\em S}urface).
  \end{itemize}
  
\item A $60^\circ$ slope, ``untilted'' pile on a smooth surface, see
  Fig.~\ref{fig:NT60-pile}. The {\em NT60} case. ({\em NT60} stands
  for {\em N}ot {\em T}ilted {\em 60} degrees slope) .
  
\item A $30^\circ$ slope, ``untilted'' pile on a smooth surface, see
  Fig.~\ref{fig:NT30-pile}, the \NTIII case. In this case we have
  studied the influence of applying a force on the lower row of discs
  by slightly displacing the corner stones.  This configuration
  (amongst others) was studied in Ref.~\cite{luding97} using
  ``spring-dashpot'' contacts and solved with molecular dynamics.

\end{itemize}

\section{The algorithm used}
\label{sec:alg}
The algorithm was implemented on a Sun computer running Maple-V
symbolic calculation software. Basically, the algorithm is looking for
a configuration of the contacts for which the solution for
the positions of the discs is compatible with all of Heaviside's
$\theta$ functions.

\begin{figure}[htb]
  \centerline{ \psfig{figure=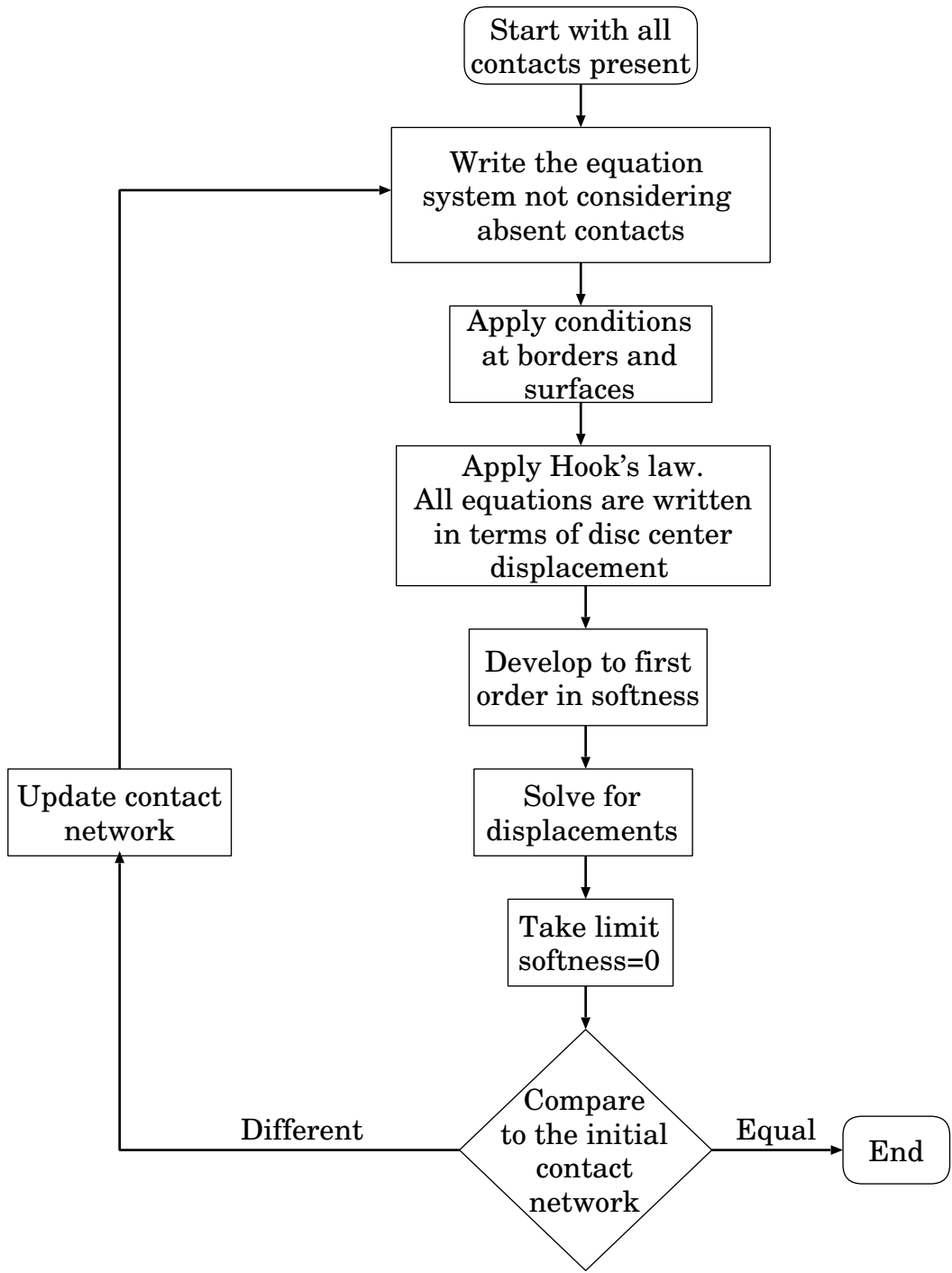,width=14cm} }
  \caption[Flowchart of the algorithm used.]
  {A simplified flowchart of the resolution algorithm used (see
  Sec.~\ref{sec:alg} for more details).}
  \label{fig:flowchart}
\end{figure}

A simplified flowchart of this algorithm is shown in
Fig.~\ref{fig:flowchart}. 
\subsection{Notations.}
\begin{figure}[htb]
  \centerline{ \psfig{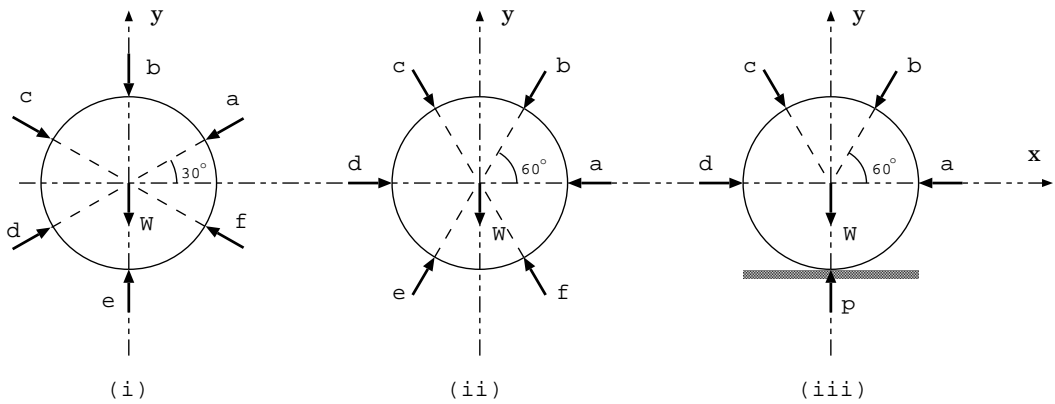} }
  \caption[Definitions of the notations used for the forces]
  {The notations used for the forces names. (i) represents the case of
    a tilted lattice ({\em TBS} and {\em TSS} cases), (ii) and (iii)
    represents the untilted one ({\em NT60} and {\em NT30}),
    where (ii) is the case of a disc in one of the upper $N-1$ layers
    and (iii) is the case of discs that touch the surface. The
    angles are those for the limit $\tau\rightarrow 0$.}
  \label{fig:force-definition}
\end{figure}

\noindent Here and below, we will use the following notations:

\begin{itemize}
\item The discs are indexed by the couple $(i,j)$ where $i$ is the row
  number, counting from the top to the bottom, and $j$ is the disc's
  position counting from the left to right. See
  Fig.~\ref{fig:TBS-pile} for example.
  
\item The forces that might act on the disc $(i,j)$ are named
  $a_{i,j}, b_{i,j}, c_{i,j}, d_{i,j}, e_{i,j}, f_{i,j}$ and
  $p_{i,j}$. Note that the notations are different for the two lattice
  orientations and that $p_{i,j}$ is only present in the untilted case
  with smooth surface (see Fig.~\ref{fig:force-definition}).
  
\item We denote with $\alpha(i_1,j_1,i_2,j_2)$ the angle between the
  horizontal and the line connecting the centers of discs $(i_1,j_1)$
  and $(i_2,j_2)$.
  
\item We designate by $x_{i,j}$ and $y_{i,j}$ the $x$ and $y$
  coordinates of the center of the disc $(i,j)$.
  
\item We call $N$ the number of layers in the pile. In the bumpy
  surface case we consider $N$ as the number of {\em free to move}
  layers.
  
\item We name $t_i$ the total number of discs on the $i$-th layer,
  $h_i$ the number of discs in one {\em half} of the $i$-th layer,
  center disc included and finally $h'_i$ the same number but
  excluding the center disc.  It is simple to see that:
    \begin{itemize}
    \item For the tilted piles (Fig.~\ref{fig:TBS-pile}
      and \ref{fig:TSS-pile}) we have $t_i=i,\quad h_i=
      \left \lfloor \frac{i+1}{2} \right \rfloor $ and $h'_i= \left
        \lfloor \frac{i}{2} \right \rfloor$.
    \item For the $60^\circ$ slope untilted lattice pile
      (Fig.~\ref{fig:NT60-pile}), $t_i=i, \quad h_i=\left \lfloor
        \frac{i+1}{2} \right \rfloor $ and $h'_i= \left \lfloor
        \frac{i}{2} \right \rfloor$.
    \item For the $30^\circ$ slope untilted lattice pile
      (Fig.~\ref{fig:NT30-pile}), $t_i= 3i-2, \quad h_i=\left \lfloor
        \frac{3i-2}{2} \right \rfloor $ and $ h'_i= \left \lfloor
        \frac{3i-1}{2} \right \rfloor $
    \end{itemize}
  \item The variable $Z$ will contain, in our algorithm, the list of
    the forces between the couples of neighboring discs which are not
    in contact. These forces will be removed from the system of
    equations.
\end{itemize}

\subsection{Resolution steps}
The general resolution steps are the following (see also the flowchart
shown in Fig.~\ref{fig:flowchart}):
\label{sec:resolution-steps}

\begin{enumerate}
\item Initialize $Z=\emptyset$. One can also start with another
  initial contact configuration closer to the solution in order to
  speed up the calculation time.
  
\item \label{item:eq-sys} Write the system of equations $\sum \vec F_i
  = 0$ for the entire pile for the $x$ and $y$ projections. In
  order to limit the size of the equation system we take immediately
  into account the $x\rightarrow -x$ symmetry of the pile. In other
  terms, we write only the projections on the $x$ axis of the equations for
  the discs $(i,j)$ with $i=1..N$ and $j=1..h'_i$ and the projections
  on the $y$ axis for $j=1..h_i$.  At the end of this step the system
  is written in terms of the forces $ a_{i,j}, b_{i,j}, c_{i,j},
  d_{i,j}, e_{i,j}$ (and eventually $p_{N,j}$) and the angles between
  the disc centers (see Fig.~\ref{fig:force-definition}).
  
\item We introduce the conditions at the border and at the bottom of
  the pile. In other terms, substitute $0$ for forces like (in the
  {\em TSS} case) $b_{i,1}, c_{i,1}, e_{N-1,j}$ etc. where $i=1..N-1,
  j=1..t_{N-1} $.
  
\item \label{item:break_contacts} We remove forces between neighbors
  not in contact. In other words, we substitute $0$ for all of the
  forces in the variable $Z$.
  
\item We apply Hook's law (equation \ref{eq:Hook}, replacing the
  $\theta$ functions by $1$), so that the forces are now expressed in
  terms of the positions of the centers of the discs.
  
\item Replace the angles by their expressions in the positions.  For
  example, in the \TSS and \TBS case, pose:
\begin{equation}
  \cos(\alpha(i,j,i-1,j))=
  \frac{x_{i-1,j}-x_{i,j}}{\sqrt{(x_{i,j}-x_{i-1,j})^2+
      (y_{i,j}-y_{i-1,j})^2}}
\end{equation}

\item At this point the equations are written entirely in terms of
  $x_{i,j}$ and $y_{i,j}$. The assumption of very high stiffness 
  enters here, we rewrite the positions in terms of the displacements
  from the lattice points $\delta x_{i,j}$ and $\delta y_{i,j}$, \ie
  we set $x_{i,j}=x^{lattice}_{i,j}+\delta x_{i,j}$ and
  $y_{i,j}=y^{lattice}_{i,j}+\delta y_{i,j}$.
  
\item We develop the equations to first order in those displacements.
  \label{item:develop}
\item We resolve the linear equations to find $\delta x_{i,j}$ and
  $\delta y_{i,j}$ . These displacements are proportional to $\tau$,
  the inverse elastic modulus.
  
\item In some cases ({\em TSS, NT60, NT30}) the system obtained in
  step~\ref{item:develop} does not have
  any solution because too many forces were removed at once at the
  previous iteration and stability can no longer be assured.  If such
  a case appears, we remove from $Z$ the force having the minimal
  value for $\left |2 R-|\vec r_{i_1,j_1}-\vec r_{i_2,j_2}|\right |$,
  in other terms we reintroduce a contact between the closest couple
  of separated discs until equilibrium is regained.
  
\item Once the displacements are known, we use Hook's law again in
  order to calculate the contact forces.
  
\item If some of those forces are found to be negative, \ie the
  corresponding contact is attractive, we add them to $Z$ (those
  forces will be eliminated in step~\ref{item:break_contacts}).  In a
  similar manner, forces currently in $Z$ which are no longer
  attractive (since for the last solution found $2 R-|\vec
  r_{i_1,j_1}-\vec r_{i_2,j_2}|\ge 0$), are removed from $Z$.
  
\item If the last step produced no change, we conclude that our
  solution satisfies all of Heaviside's $\theta$ functions and the
  algorithm stops, returning the forces, $Z$ and the displacements.
  As we mentioned before this proves the existence of a solution, but
  there might be others.  Otherwise, the list $Z$ is updated and the
  algorithm returns to step~\ref{item:eq-sys}.

\end{enumerate}
\subsection{Computer resources used}
The computer time needed in order to get the final force configuration
varied from some minutes (4 layers) to several days (more than 20
layers) on a Sparc 20 station. The maximum size varied from one
configuration to the other, because some (like the {\em TSS} or {\em
  NT30}) needed more iterations than other cases before arriving to
the correct contact configuration.  Another limitation is memory,
large systems create huge systems of equation that can occupy large
amount of memory (we used up to 30Mb of RAM in some cases).

\section{Results}
\label{sec:results}
\noindent For each of the geometrical configurations we present several
interesting characteristics derived from the solution:
\begin{description}
\item[The force and contact networks:] When the algorithm
  stops, we obtain a contact network and its superposed force
  network. We will represent those networks in a single plot where a
  dashed line represents an absent contact, a full line represents a
  contact and it is drawn with a width proportional to the amplitude
  of the force. In some cases we observe the existence of a limit case
  where a contact exist but the force vanishes (osculatory discs).
  This kind of contacts will be represented by a dash-dot line. 
  These plots will give us an idea of the effective lattice in
  different zones of the pile and the possible existence of an arching
  effect. 
\item[The stress tensor:] The stress tensor $\sigma_{\alpha\beta}$,
  averaged over one particle and in the case of point-like contacts is
  given by:
  \begin{equation}
    \sigma_{\alpha\beta}=(1/V)\sum_k F_\alpha^k r_\beta^k
    \label{eq:stress}
  \end{equation} 
  where the sum runs over all external forces acting on the particle,
  $F_\alpha^k$ is the $\alpha$-th component of the the $k$-th force,
  $r_\beta^k$ is the $\beta$-th component of its contact point
  position and $V=\pi R^2$ the volume of each particle. Since we neglect
  tangential forces the particles are torque free and the stress
  tensor is symmetric.  In these plots, the stress tensor for each
  particle will be represented in by two segments of length
  proportional to the the eigenvalues and pointing at the direction of
  the eigenvectors (principal and secondary axis).  The stress tensor
  will help us compare our results to results from various continuum
  mechanics approaches, especially to the fixed principal axis
  hypothesis.
\item[Pressure profile on the base of the pile:] The normal and (when
  present) horizontal forces applied by the pile on the supporting
  surface will be plotted as a function of the position $x$. These
  plots will provide a comparison of our results to the experimental
  data, mainly in order to verify whether a dip is present or not.
\item[Size dependency of the forces:]
  By solving for different pile sizes we can trace the evolution of the
  forces acting on a given disc as a function of the size of the
  pile. In order to get the maximum number of data points we
  concentrated on disc $(2,1)$. 
\end{description}

\subsection{``Bumpy'' surface tilted lattice : {\em TBS}}
In this section we present the results for a piling of discs posed on
2 layers of discs attached to the lattice, as shown in
Fig.~\ref{fig:TBS-pile}.
\subsubsection{The resulting force network}
In this case one can notice that the missing contacts tend to appear
on the flanks of the pile, while the in the inner part of the pile all
neighboring discs are in contact. This behavior is incompatible with
the basic \FPA assumption since a particle at the open surface is in a
completely different situation once it is buried.  None of the contacts
downwards are missing and we can notice that these contacts are the
prefered paths for the forces thus excluding any arching.
\subsubsection{The stress tensor}
The resulting stress tensor is shown in Fig.~\ref{fig:tensor-TBS}.  As
we could expect, we do not observe in this geometry a fixed principal
axis direction, but rather a typical result that would follow from the
traditional IFE (Incipient Failure Everywhere) assumption
(Ref.~\cite{wittmer97}, Sec.~2.5). The stress tensors' variations are quite
smooth and all of the pile seems to have the same behavior. \rem{!!!!!}
\subsubsection{The pressure profile}
In the {\em TBS} case it is not straightforward to define uniquely the
pressure or the normal force distribution on the supporting surface
since it is in contact with two different layers. We show in
Fig.~\ref{fig:pressure-TBS} the normal and shear forces on the last
layer of discs, ignoring those on the layer $N-1$. Because we took
into account only the $N$-th layer the sum of the normal forces is
smaller than the total weight of the pile.  Anyhow, it is clear that
the
distribution does not show a dip but rather a maximum%
\footnote{Both
  distributions of normal forces on the $N$-th and $N-1$ layer are
  peaked under the apex of the pile.}.
\subsubsection{Dependence on the size of the pile}
The evolution of the forces acting on the disc $(2,1)$ is plotted in
Fig.~\ref{fig:dependence_in_size-TBS}.  For small values of $N$, the
finite size effects are large and the forces fluctuate considerably.
But as the size increases, these effects disappear, and a linear
regime is established.  The important point here is that all the
forces, even those at the top of the pile are highly correlated to the
size and ``feel'' each added layer.  We can expect that this should
not be the case when the size of the pile gets even bigger and a
saturation should occur. The question is how will it occur? Will
asymptotically the forces tend to a constant value, or will they
rather continue changing linearly until one of them (in the figure's
case $e_{2,1}$) will reach zero causing the corresponding contact to
disappear and the other forces to stabilize?  We are not able to study
the behavior of the forces at the point where, for example, $e_{2,1}$
vanishes, since the computer resources needed would be too large for
the corresponding pile sizes\footnote{For 22 layers pile in this case,
  the calculation needed about 30Mb of RAM and took more than 3 days.}
(more then 30 layers).

\subsection{Tilted lattice, smooth surface : {\em TSS}}
The geometrical situation is presented in Fig.~\ref{fig:TSS-pile}.
Discs in the last row are free to move without friction on the
surface, only the corner stones (in gray) are fixed on the lattice and
are not allowed to move.  This situation is technically harder than
the previous one because on a bumpy surface the algorithm only removed
contacts in every iteration without ever creating them again in the
following iterations, so that every iteration approaches the final
solution. Here, this is not the case. Contacts often reappear, making
the approach to the final solution much slower.
\subsubsection{Force network}
\label{sec:network-TSS}
An example of the force network obtained is shown in
Fig.~\ref{fig:network-TSS}. This network differs considerably from the
one in the \TBS case. First of all the number of absent contacts
is much larger and they appear mainly in the center of the pile and
not just on the surface as in the \TBS case. Secondly, a new case
has appeared for which the force vanishes exactly, even though the
corresponding contact does exist (osculatory discs). Again, nearly all
of the vertical contacts are present so no arching is observed.
\subsubsection{Stress tensor}
\begin{figure}[htb]
  \centerline{\psfig{figure=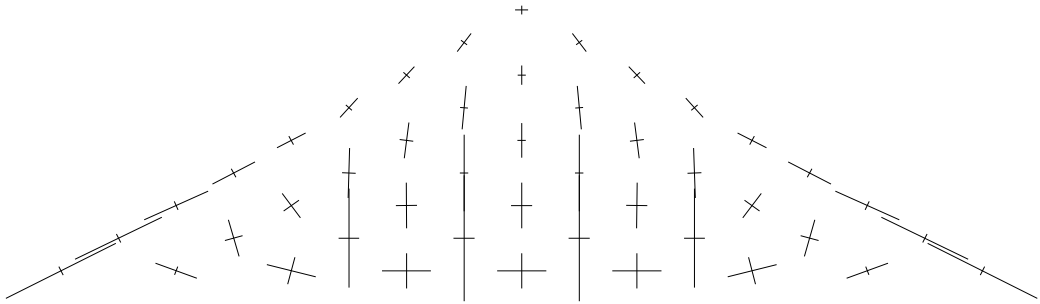,width=14cm}}
  \caption[Principal and secondary axis of the stress
  tensor in a pile of 10 layers for the {\em TSS} case.]{Principal and
    secondary axis of the stress tensor in a pile of 10 layers for the
    {\em TSS} case (see Sec.~\ref{sec:results} for more details).}
  \label{fig:tensor-TSS}
\end{figure}

The stress tensors' principal and secondary axis are shown in
Fig.~\ref{fig:tensor-TSS}. Again, the results differ from the \TBS
case. \rem{!!!} We can observe 3 regions with different behaviors with
an abrupt transition:

\begin{itemize}
\item Discs on the surface of the pile, having the principal axis
  pointing along the surface.
\item Discs in the inner part of the pile, having the principal and
  secondary axis pointing horizontally or vertically.
\item Discs at the shoulders of the pile with no clear prefered
  direction.
\end{itemize}
In this case our result is in contradiction to the basic assumption  
of the {\em FPA} model since the stress tensor on the surface of
the pile differs considerably from the one of a buried particle.
\subsubsection{Pressure profile}
\begin{figure}[htb]
  \centerline{
    \psfig{figure=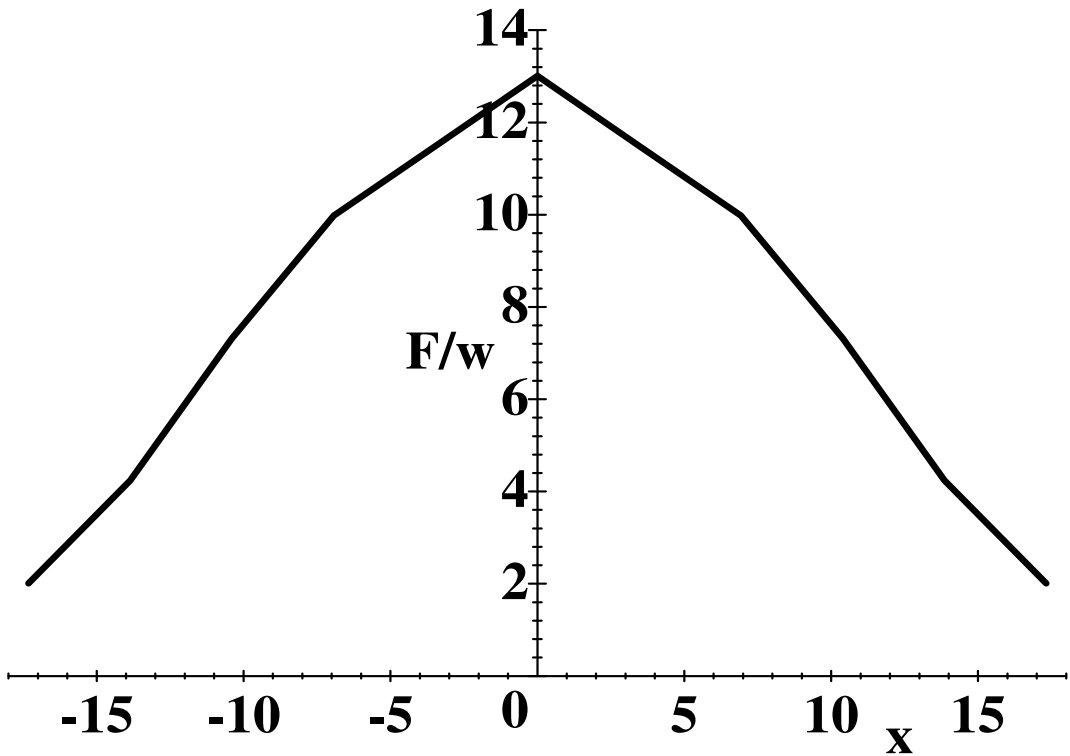,width=14cm,height=10cm}}
 \caption[Pressure profile on the base of a 13 layers {\em TSS} pile.]%
 {Pressure profile on the base of a 13 layers {\em TSS} pile (see
   Sec.~\ref{sec:results} for more details).}
 \label{fig:pressure-TSS}
\end{figure}

The normal forces acting on the surface of a 13 layer pile%
\footnote{The biggest pile size we were able to calculate for this
  case.}  are shown in Fig.~\ref{fig:pressure-TSS}. The shear force
acting on the bottom layer is zero because of the smoothness of the
surface.  This pressure profile is close to the one in the {\em TBS}
case displayed in Fig.~\ref{fig:pressure-TBS}. As in the {\em TBS}
case we do not observe a dip but rather a maximum.
\subsubsection{Variations of the forces with the size of the pile}
\begin{figure}[htb]
  \centerline{
    \psfig{figure=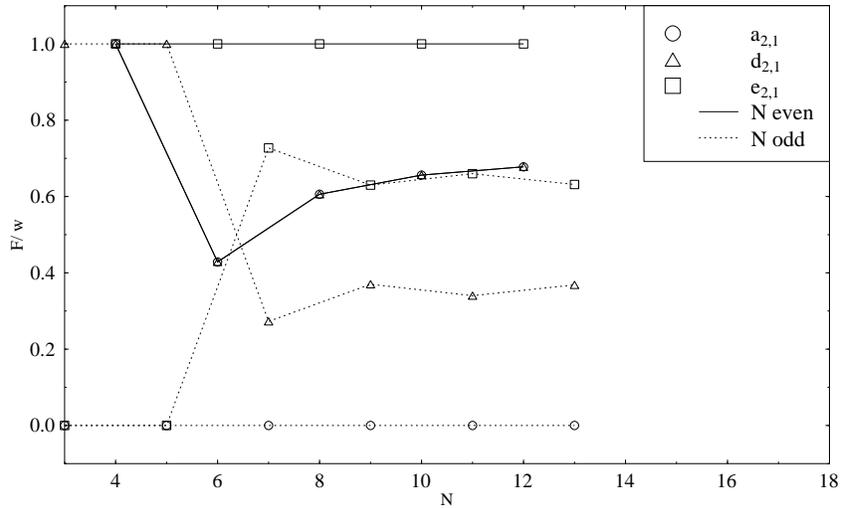,width=14cm}}
  \caption[Variations of forces with the size of the pile in the {\em
    TSS} case]{Variation of the different forces acting on the disc
    $(2,1)$ as a function of the size of the pile measured in layers; $N$,
    in the {\em TSS} case. For clarity $f_{2,1}$ is not shown since we
    obtain for all sizes that $f_{2,1}=1-e_{2,1}$. Notice the
    dependence in the parity of $N$ that stays pronounced even for
    large $N$ (see Sec.~\ref{sec:results} for more details).}
  \label{fig:dependence_in_size-TSS}
\end{figure}

The variations of forces acting on disc $(2,1)$ are presented in
Fig.~\ref{fig:dependence_in_size-TSS} (compare with
Fig.~\ref{fig:dependence_in_size-TBS}). We observe a strong dependence
on the parity of $N$ and a saturation of the forces for large pile
sizes of the same parity. We do not have enough data in order to
determine whether the dependence on the parity of $N$ continues for
even larger piles.

\subsection{The \NTVI case: ``untilted'' lattice, $60^\circ$ slope pile}
The system in this case is shown in Fig.~\ref{fig:NT60-pile}. When one
does not consider the horizontal contacts (equivalent to taking a
slope slightly inferior to $60^\circ$) it is easy to calculate
analytically the forces and find that the distribution of normal
forces is uniform on the base and that the shear force on the base
varies linearly with the position~\cite{hong93,huntley93,menon}. In
this case we always get a contact network in which {\em all}
horizontal contacts are absent, thus giving the same solution as the
one in Ref.~\cite{hong93}.  This gives us another verification for our
method and an indication about the uniqueness of the solution.
Numerical calculations give the same effect~\cite{luding97}.

\subsection{The \NTIII case: ``untilted'' lattice, $30^\circ$ slope pile}
The configuration in this case is the one shown in
Fig.~\ref{fig:NT30-pile}. Like in the {\em TSS} case, we fix the
corner stones in order to keep the heap stable.  This particular
geometry was proposed (amongst others, not studied here)
in Ref.~\cite{luding97}, and studied by numerical simulation using the
molecular dynamics method.  Since in this kind of pile the number of
discs grows faster with the number of layers than in the previous
configurations, we were only able to calculate the force network for
piles having less than 7 layers, making a systematic size effect study
impossible.

\subsubsection{The force network}
\begin{figure}[htb]
  \centerline{ \psfig{figure=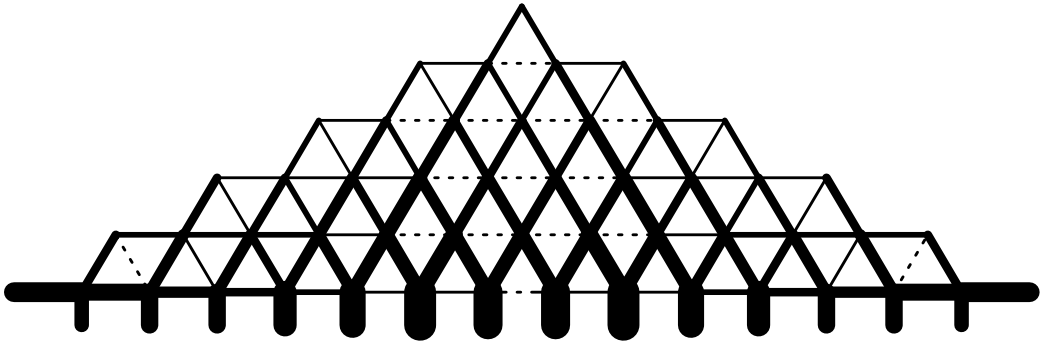,width=14cm}}
  \caption[Force network in the {\em  NT30} case for a 6 layer
  pile.]{Force network in the {\em  NT30} case for a 6 layer pile (see
  Sec.~\ref{sec:results} for more details).}
  \label{fig:network-NT30}
\end{figure}

\begin{figure}[htb]
  \centerline{ \psfig{figure=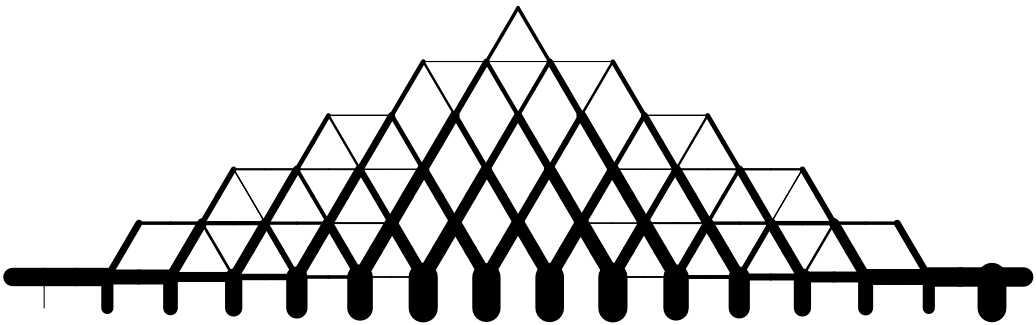,angle=270,width=14cm} }
  \caption{The force network obtained for the {\em NT30} case  by
    molecular dynamics simulation of a pile of 6 layers {\protect
      \cite{luding-private}}. }
  \label{fig:network-MD}
\end{figure}

An example of the force network obtained in this case is shown in
Fig.~\ref{fig:network-NT30}. We notice 3 regions somewhat similar to
those observed in the \TSS case (see Sec.~\ref{sec:network-TSS} and
Fig.~\ref{fig:network-TSS}). Only at the outermost parts of the pile
traces of arching are seen where some contacts downwards are absent.
In the central part of the pile all of the horizontal contacts are
absent, and between those two zones all contacts are present. In
Fig.~\ref{fig:network-MD} we show numerical results obtained by
molecular dynamics and provided by
Luding~\cite{luding-private,luding97} for the same pile. They show
excellent agreement with ours, even though the numerical results
present some differences (some contacts that are absent and should be
present or the vice versa) and imperfections (notice that the
numerical force network is not exactly symmetric especially at the
corner stones).
\subsubsection{The stress tensor}
\begin{figure}[htb]
  \centerline{ \psfig{figure=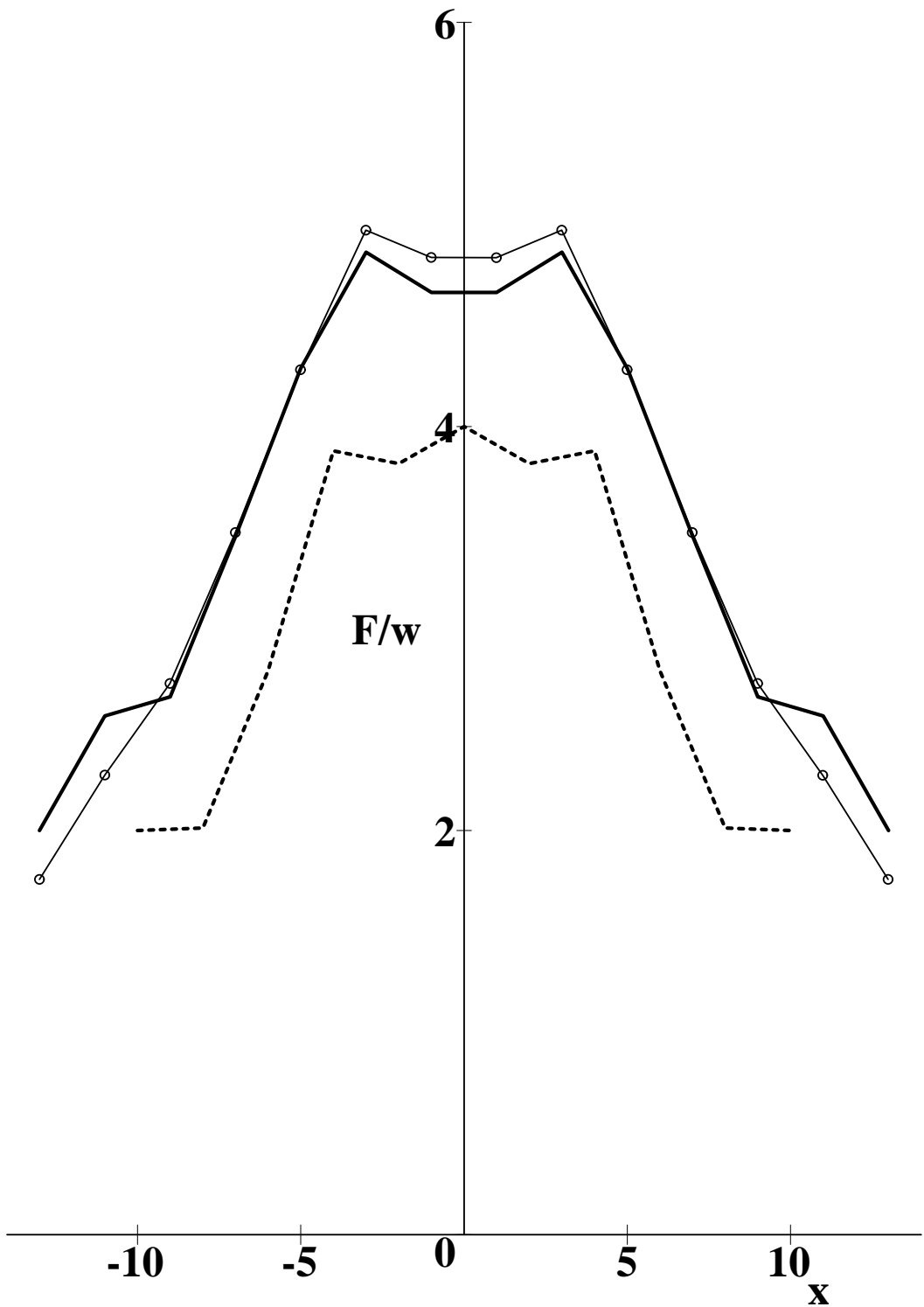,width=14cm}}
  \caption[The principal and secondary axis in the {\em  NT30} case for
  a 6 layer pile (in the cross representation) and a comparison to
  results obtained in Ref.~{\protect \cite{luding97}}]{The principal
    and secondary axis in the {\em NT30} case for a 6 layer pile (in
    the cross representation) and a comparison to results obtained in
    Ref.~{\protect \cite{luding97}} (represented by ellipses). Notice
    the excellent agreement between the two (see
    Sec.~\ref{sec:results} for more details).}
  \label{fig:tensor-NT30}
\end{figure}

The stress tensor in this case, shown in Fig.~\ref{fig:tensor-NT30},
is quite similar to the one obtained for the \TBS pile (see
Fig.~\ref{fig:tensor-TBS}), and no \FPA is observed. For comparison,
results from Ref.~\cite{luding-private,luding97} are superposed 
(represented by ellipses). The agreement is so good that the plots are
hardly distinguishable.
\subsubsection{The pressure profile}
\begin{figure}[htb]
  \centerline{
    \psfig{figure=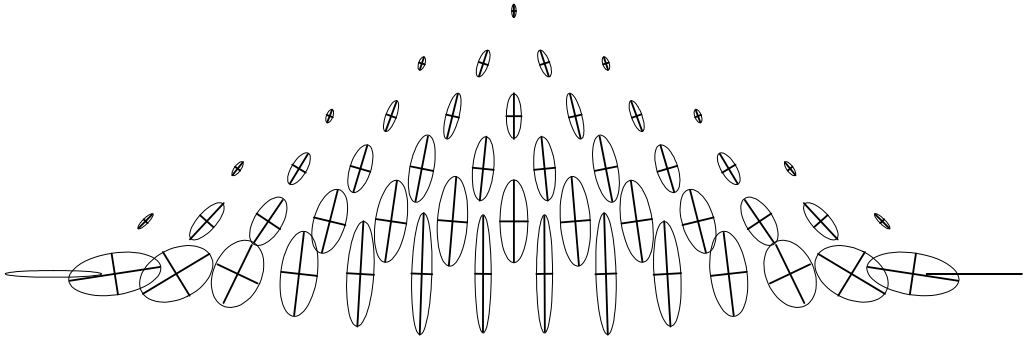,width=14cm}}
     \caption[The normal pressure profile for the {\em NT30} case]
     {The normal pressure profile for the {\em NT30} case for a 5
       (dashed line) and 6 (full line) layer pile. For the 6 layer
       pile we obtain a small dip below the apex of the pile (see
       Sec.~\ref{sec:results} for more details).}
  \label{fig:pressure-NT30}
\end{figure}

The pressure profiles for a 5 layer pile (dashed line) and a 6 layer
pile (full line) are shown in Fig.~\ref{fig:pressure-NT30}.  Both
present a plateau and a shallow minimum under the apex of the pile or
near it. We also observe a small depression at the shoulders of the
pile that, as we will see in Sec.~\ref{sec:push}, will play an
important role when the corner stones are pushed in. We also plot on
the same figure results from Ref.~\cite{luding-private,luding97} (in
circles) that, again, agree well with ours.
\subsubsection{The effect of displacing the corner stones}
\label{sec:push}
Another point studied in Ref.~\cite{luding97} is the effect of
horizontally pushing or pulling the corner stones on the bottom layer.
It was observed that pushing in those discs, \ie fixing their position
on the a position slightly closer to the axis of the pile, causes the
pressure distribution on the base to present a deeper dip than the one
observed without it.

We studied the same problem using an exact symbolic algorithm which is
an extension of the previous one. Since all calculations are done to
the first order in $\tau$ we should only consider displacements
proportional to $\tau$ \ie $\delta x_{N,1}=q\tau$ and $\delta
x_{N,t_N}=-q\tau$, otherwise the resulting forces will be infinite.

To find the correct contact configuration for a given value of the
parameter $q$, namely $q_{end}$ we proceed in the following way:

\begin{description}
\item{1.} We apply the algorithm described in
  Sec.~\ref{sec:resolution-steps} in order to solve for $q=0$.
\item{2.} For the contact network obtained in step 1 and for an
  arbitrary $q$ we find the distance between every couple of neighboring
  discs : $d_{i,j}(q)$ to first order in $\tau$. Those distances
  depend on $q$ linearly, so we can write :
  $d_{i,j}(q)=\alpha_{i,j} q+\beta_{i,j}$.
\item{3.} We solve all the equations $d_{i,j}(q)=2R$ for
  $q$. When a solution exists it is unique, we denote it $q'_{i,j}$.
  In other terms, $q'_{i,j}$ are the values of $q$ where the contact
  network might change, \ie contacts might disappear or reappear.
\item{4.} We calculate; 
\begin{equation}
  q''=\left\{
    \begin{array}{ll}
      \min (q'_{i,j} | (q'_{i,j}>q) & \hbox{if } q_{end}>0 \\
      \max (q'_{i,j} | q'_{i,j}<q) & \hbox{if } q_{end}<0
    \end{array}
  \right.
\end{equation}
In practice, we find the next $q$ for which the contact network
might change.
\item{5.} If $|q''|>|q_{end}|$ then no change in the contact network
  will happen between $q$ and $q''$ and since the network and
  $d_{i,j}(q)$ are known for this interval of values, and hence for
  $q=q_{end}$, we can calculate the forces for $q=q_{end}$ easily from
  equation (\ref{eq:Hook}) and the algorithm stops.
\end{description}
The expression of each force, as a function of the positions, is
continuous around $q''$ (equation~\ref{eq:Hook}) so that no
discontinuity can occur in the forces if contacts having
$d_{i,j}(q'')=2R$ disappear, only $\alpha_{i,j}=d d_{i,j}(q)/dq$ might
be modified.
\begin{description}
\item{6.} We modify the contact network to take into account the
  changes supposed to take place at $q''$. We set $q=q''$ and jump
  back to step~2.
\item{7.} In some cases the modification we applied to the contact
  network must be reviewed since in certain cases a force $F_{i,j}$
  vanished in $q''$ but $\hbox{sign}(\alpha_{i.j}(|q|<|q''|))=
  -\hbox{sign}(\alpha_{i.j}(|q|>|q''|))$ so that the corresponding
  contact should not change. For example, two neighboring discs that
  were approaching one another for 
  $q<q''$ \ie $\left.\alpha_{i,j}\right|_{q\rightarrow
    q''^+}<0$ and for which $d_{i,j}(q'')=2R$ will be supposed in
  contact in the
  new network, but if $\left.\alpha_{i,j}\right|_{q\rightarrow
    q''^-}>0$ this contact should not reappear. When the algorithm
  detects such cases the contact network is corrected.
\end{description}

\begin{figure}[htb]

\centerline{\psfig{figure=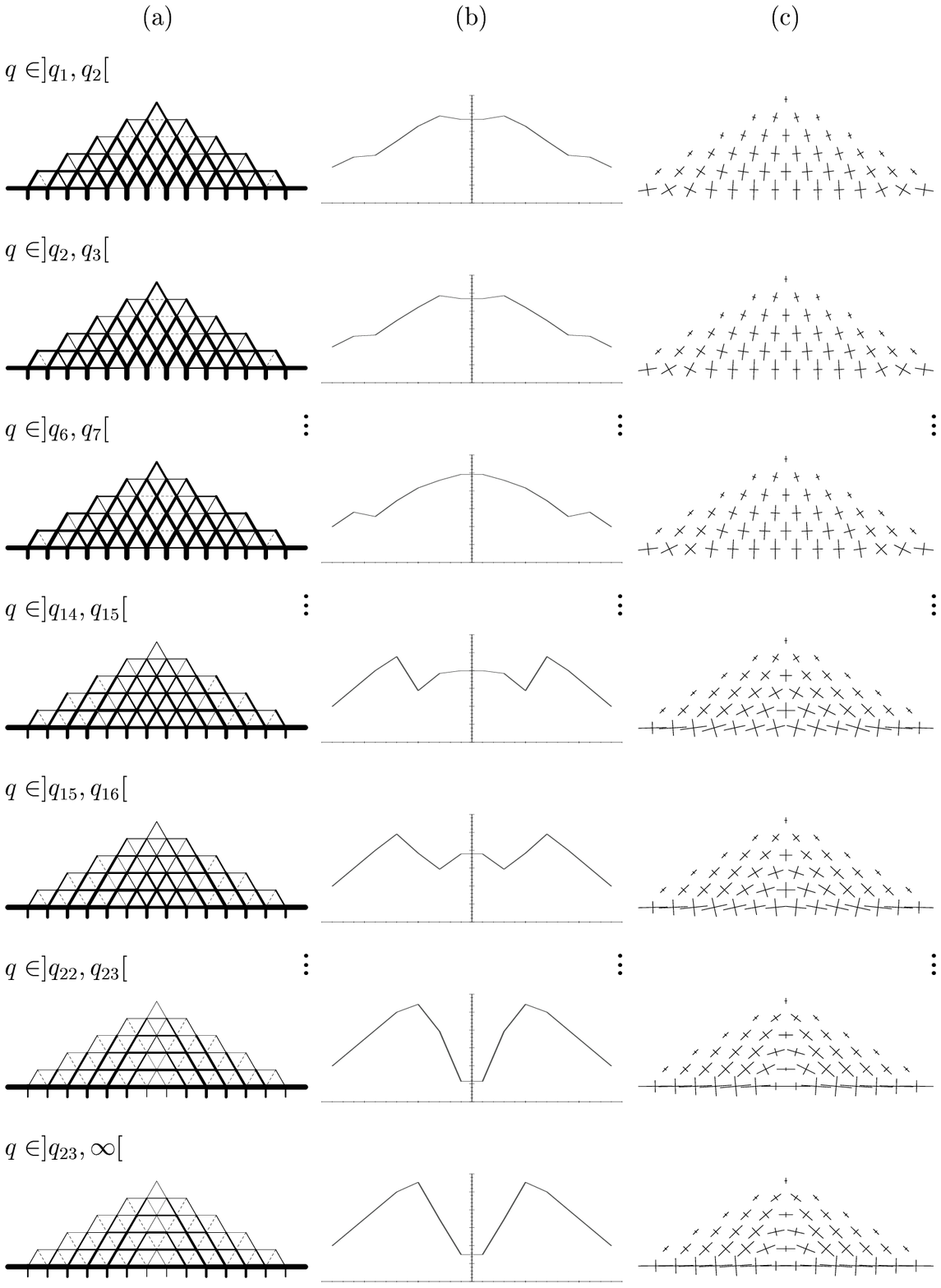,width=\linewidth}}
  \caption[The changes in the force network, pressure profile and the
  stress tensor for growing values of $-q$]{The changes in the force
    network (a), pressure profile (b) and the stress tensor (c) for
    growing values of $-q$ (see table (\ref{tab:qlist}) for the values
    of $q_i$) for a 6 layer {\em NT30} pile. In column (b) the
    abscissa represents the $x$ axis, and varies between $-14$ and
    $14$ in disc radii unit, and the ordinate represents the force
    measured in disc weights $F/w$ varying in the range $0$ to $6$.
    The length of the representation of the principal axis of the
    stress tensor (c) and the width of the lines in the force network
    plots (a) are in a logarithmic scale.}
  \label{fig:NT30-push-seq}
\end{figure}

\begin{figure}[htb]
  \centerline{\psfig{figure=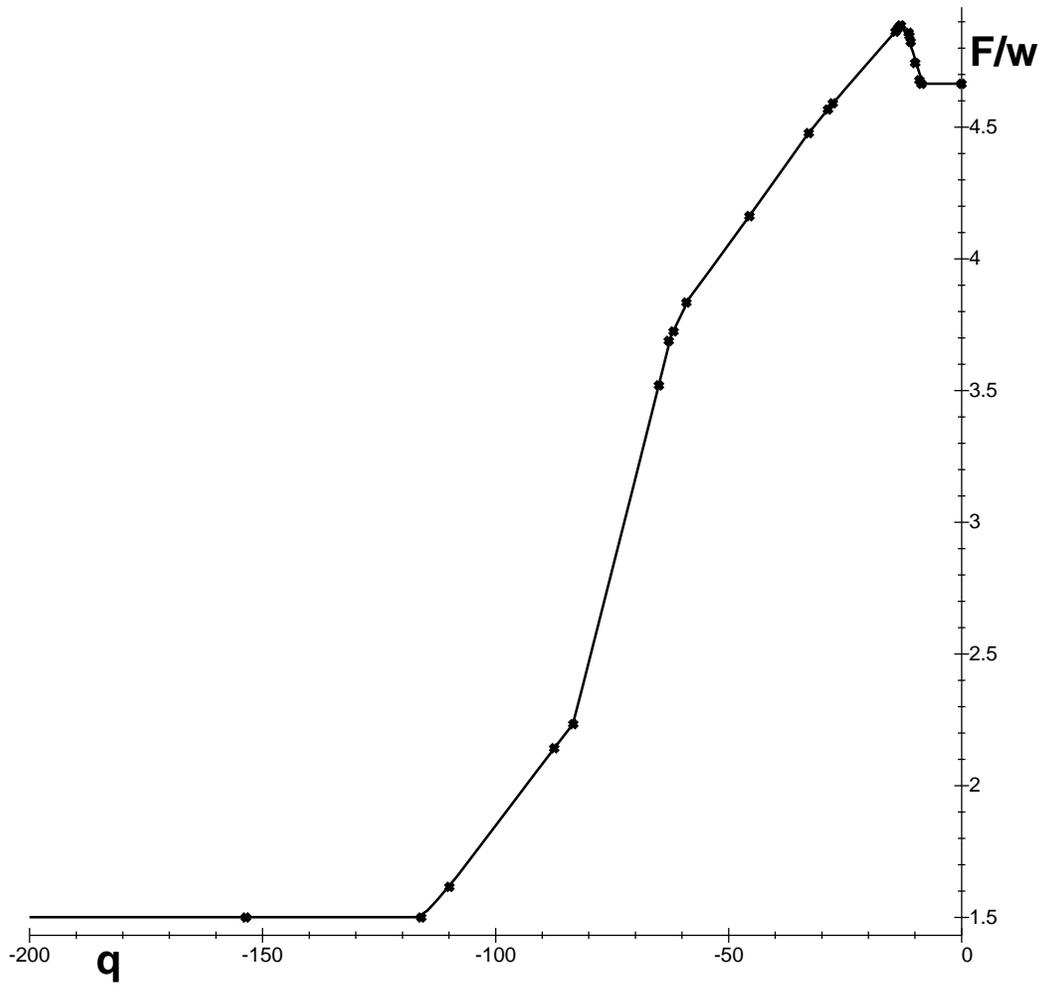,width=14cm}}
  \caption[Variation of the pressure under the apex of the pile with
  $q$ in the {\em NT30} case]{Variation of the pressure under the apex
    of the pile with $q$ in the {\em NT30} case. The dots represent
    the values of $q$ where the contact network rearranges (see table
    \ref{tab:qlist}).}
\label{fig:NT30-dip}
\end{figure}

An example for results of the implementation of this method is shown in
Fig.~\ref{fig:NT30-push-seq} and the variations of the normal forces
on the supporting surface are shown in Fig.~\ref{fig:NT30-dip}. The
values of $q$ where the contact network rearranges for the 6 layer
pile are given in table~\ref{tab:qlist}. We observe several interesting
points in this case.  The initial pile is insensitive to positive
values of $q$, in other terms no change in the force network appears
when the corner stones are pulled apart. This appears to be a
characteristic of the solution for all lattice orientations when the
surface is smooth (not exposed here).  On the contrary, when pushing
the corner stones together $(q<0)$ the network is restructured at
many values of $q$ (table~\ref{tab:qlist}). The number of those values
is finite so that for large values of $-q$ the contact network no
longer changes. At this point (see Fig.~\ref{fig:NT30-push-seq}) the
network is characterized by a large number of absent vertical contacts
and all of horizontally neighboring discs in contact, which is exactly
what one might classify as an arching effect.  As observed
in Ref.~\cite{luding97} 
the arching is accompanied by the appearance of a
pronounced depression under the apex and a {\em FPA} situation.  We
were able to calculate the exact evolution of the pressure under the
apex of the pile as a function of $q$, see Fig.~\ref{fig:NT30-dip}.
For large values of $-q$ the pressure under the apex of the pile is
decreasing with $-q$ until it reaches its asymptotic value.
Surprisingly, we also observe a region for small values of $-q$ for
which the minimum is less pronounced and even becomes a maximum (see
the third frame in Fig.~\ref{fig:NT30-push-seq}).  The mechanism
of the this change can be understood from
Fig.~\ref{fig:NT30-push-seq}. It is not the small depression under
the apex of the pile, observed for $q=0$, that is the source for the dip
when $-q$ is big, but rather the depressions observed at the flanks of
the pile. These depressions move toward the center to create the dip.

\section{Conclusions}
\label{sec:conclusions}
In this paper we have presented an approach capable of producing exact
results for the force and contact networks for piles of regularly
packed hard discs for a large variety of cases: different
orientations, geometries and supporting surfaces. The advantage of
this method is that results are completely reliable since they a free
of roundoff or other numerical errors. We obtain a wide variety of
results for the different realizations we study. We conclude that
lattice orientation and the characteristics of the supporting surface
have a very important impact on the physical properties of the pile
even when the pile is globally the same, and the contacts are of the
same nature.  Permitting propagation of forces vertically downwards
when using a ``tilted'' lattice (\TBS and \TSS cases) leads to a
behavior completely contrary to arching where the contact forces of
the largest amplitude are vertical. On the other hand, when no
vertical propagation is possible arching is more likely to occur, like
we notice in the ``untilted'' case \NTIII.  Arching can also be
stimulated by imposing external constraints on the system that favor
the horizontal contacts, like pushing in the corner stones.  We were
able to follow step by step the reorganization of the contact network
while the pile is increasingly controlled by arching and a dip is
established as the constraints grow.  Experimental evidence for the
importance of the orientation of the lattice of 2D granular systems
was already observed for dynamic systems~\cite{duran94b}, but is
generally ignored for static systems where the ``untilted'' lattice is
the one generally used.  Since real granular systems are disordered,
real effects should not depend on the local orientation of the
lattice. Much care should be taken when interpreting results that do
depend on the orientation.

\begin{figure}[htb]
  \centerline{
    \psfig{figure=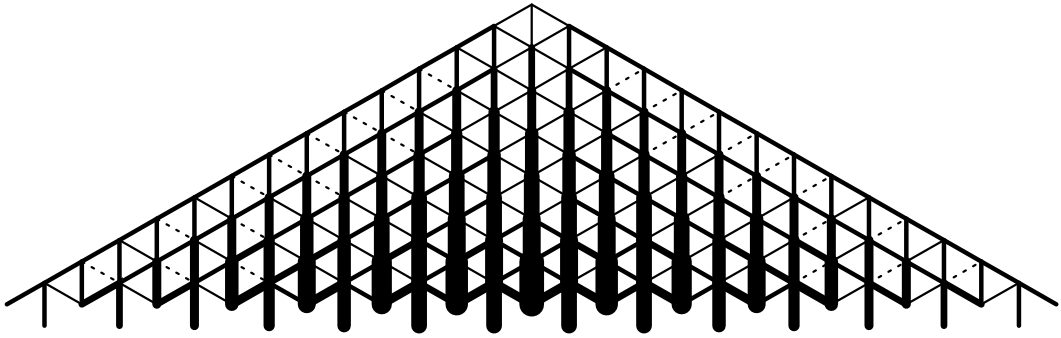,width=14cm} }
  \caption[The force network for a pile of 15 layers in the {\em TBS}
  case]{The force network for a pile of 15 layers in the {\em TBS}
    case. The line width is proportional to the force amplitude, the
    dashed lines represent absent contacts (see Sec.~\ref{sec:results}
    for more details). }
  \label{fig:network-TBS}
\end{figure}

\begin{figure}[htb]
  \centerline{ \psfig{figure=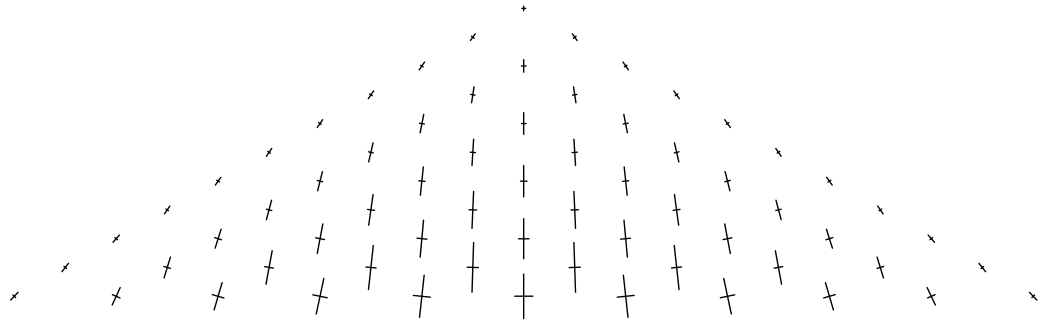,width=14cm}}
  \caption[The principal and secondary axis of the stress tensor for a
  pile of 11 layers for the {\em TBS} case.]{The principal and
    secondary axis of the stress tensor for a pile of 11 layers for
    the {\em TBS} case (see Sec.~\ref{sec:results} for more
    details).}
  \label{fig:tensor-TBS}
\end{figure}

\begin{figure}[htb]
  \centerline{ \psfig{figure=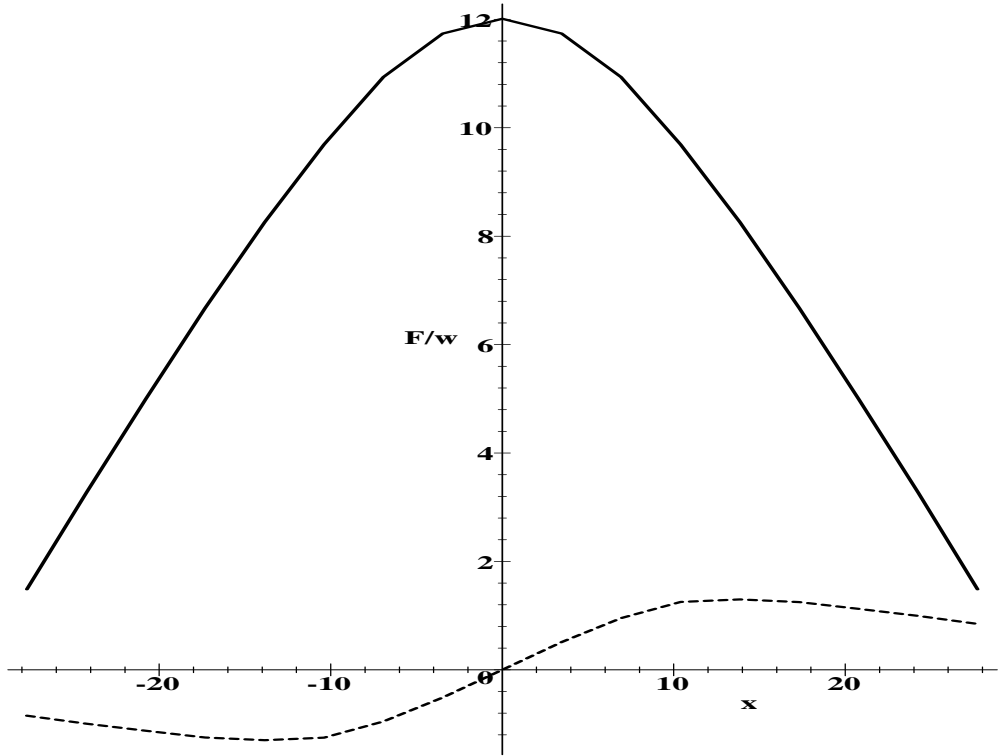,width=14cm,height=12cm}}
  \caption[Forces acting on the base of the pile in the {\em TBS}
  case]{The normal (full line) and shear (dashed line) forces on the
    base of a pile of 17 layers in the {\em TBS} case. We do not
    observe any dip in the normal force distribution. The shear forces
    vanish at the center of the pile as expected from symmetry (see
    Sec.~\ref{sec:results} for more details).}
  \label{fig:pressure-TBS}
\end{figure}

\begin{figure}[htb]
  \centerline{
    \psfig{figure=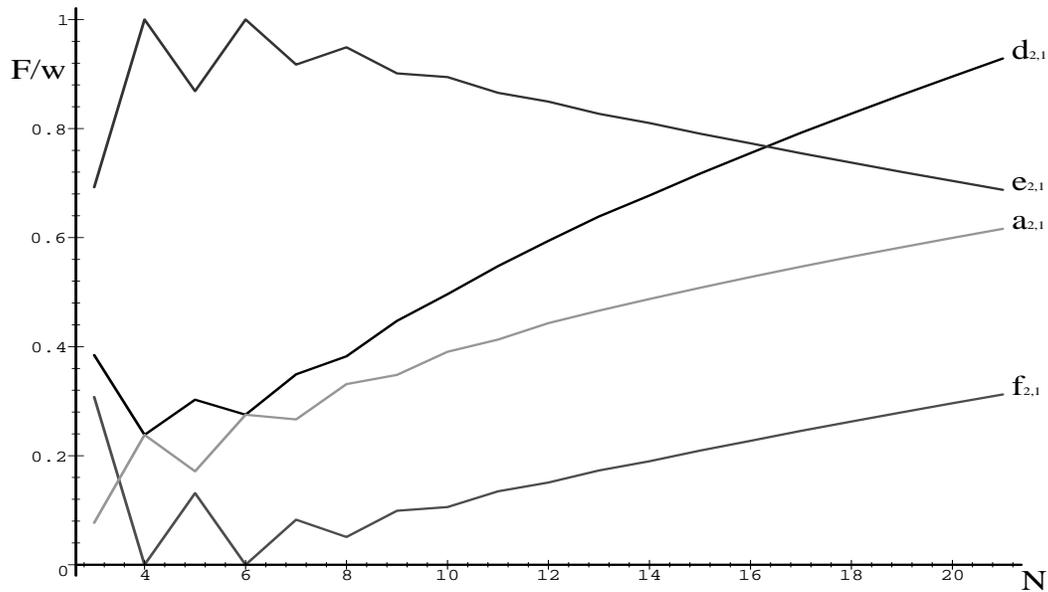,width=14cm,height=8cm} }
  \caption[Variation of the forces with the size of the pile in the
  {\em TBS} case]{The dependence of the forces acting on disc (2,1) on
    the size of the pile for the {\em TBS} case. $N$ is the number of
    layers, a,d,e and f are forces acting on the disc as defined in
    figure \ref{fig:force-definition}. All forces are normalized
    by a disc weight (see Sec.~\ref{sec:results} for more details).}
  \label{fig:dependence_in_size-TBS}
\end{figure}

\begin{figure}[htb]
  \centerline{ \psfig{figure=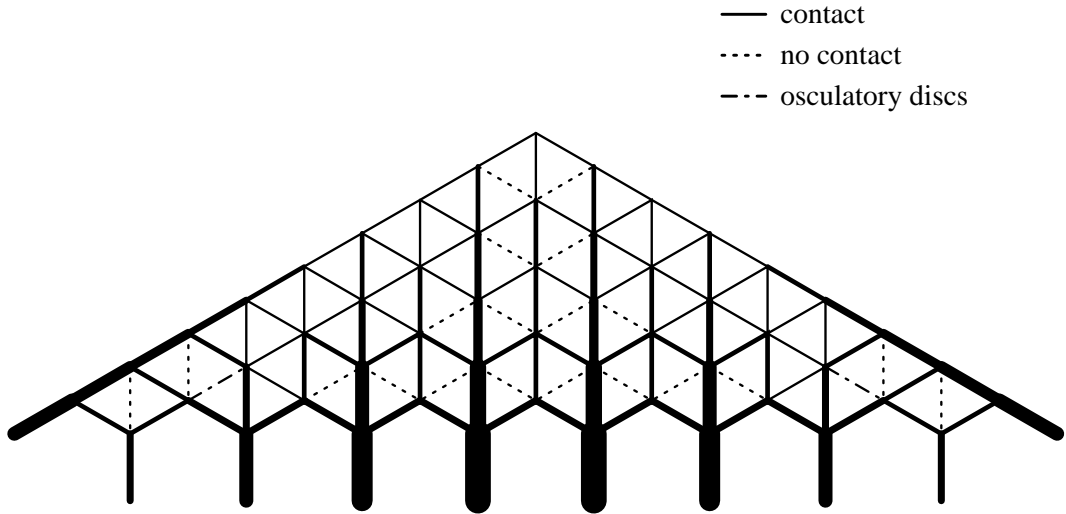,width=14cm}}
 \caption[The force network of a pile of 10 layers in the {\em TSS}
 case.]{The force network of a pile of 10 layers in the {\em TSS} case
   (see Sec.~\ref{sec:results} for more details).}
 \label{fig:network-TSS}
\end{figure}

We confirm the claims in Ref.~\cite{savage97} about the sensitivity of
the results to the boundary conditions at the base of the pile,
especially the roughness of the surface has a large influence on both
the contact network and the stress tensor (compare
Fig.~\ref{fig:network-TBS} to~\ref{fig:network-TSS}
and~\ref{fig:tensor-TBS} to~\ref{fig:tensor-TSS}).

 The results we get with this method can be used to check the
precision of widely used numerical simulations, for much larger
systems. We were able to do so in the {\em NT30} case producing the
same results as in Ref.~\cite{luding97}. We also confirm the effect
that pushing the corner stones have on the pressure profile \ie
the dip becoming pronounced. 

On the other hand, we do not get
comparable results when pulling {\em out} the corner
stones. In Ref.~\cite{luding97} asymmetric solutions were obtained in this
case, and might indicate the existence of more than one solution or
of a numerical flaw. We do not observe any rearrangement of the contacts
when the corner stones are pulled apart, which seems to be a
characteristic of all solutions with $q=0$. One must keep in mind,
however, that in Ref.~\cite{luding97} results are given for a finite value
of the stiffness, while here the stiffness is infinite.

In the \NTIII case, (Fig.~\ref{fig:pressure-NT30}) the pressure on the
base of the pile presents, for an even number of layers, the famous
dip observed under the apex of granular heaps~\cite{smid81,huntley97}.

We also observe, as in Ref.~\cite{luding97} that the dip becomes more
pronounced as we increase the applied force on the corner stones, but
we also conclude that this dip is not formed by deepening of the
shallow minimum observed for $q=0$ but rather from the small
depressions at the flanks that move toward the center of the pile when
$-q$ increases.

However, in light of the discussion above, concerning the sensitivity
of results to the type of supporting surface and lattice orientation,
and since a dip was {\em not} observed for the ``tilted'' piles
(Fig.~\ref{fig:pressure-TBS} and~\ref{fig:pressure-TSS}), we believe
that one cannot conclude that this simple model contains the
physical ingredients that give rise to the depression under the apex
of the pile or to the \FPA situation.

The method we used can be easily adapted to other geometries like
silos or 3D pilings of spheres. One might also consider the
introduction of other physical ingredients like friction or
polydisperse disc sizes.  The introduction of friction doubles the
number of degrees of freedom of the system, since tangential
components of the contact forces are added to each of the contact
forces. On the other hand Coulomb's friction law does only supply us
with an inequality $ F_{||}\leq\mu F_\perp $ rather than an equality.
One can face this problem in two ways:

\begin{itemize}
\item Use the inequality solving facility in {\em Maple} in order to
  get bounds to the force values. This would probably be a
  difficult task mainly because of the $\theta$ functions.
\item Introduce a supplementary assumption about the friction forces
  like $ F_{||}=\mu F_\perp $ which is the {\em IFE} assumption, or
  any other relation between the forces.
\end{itemize} 

The introduction of disorder into the system (polydisperse grains) does
not present conceptual problems and could be faced in principle in the
framework of our approach. However, it would be hard , in this case,
to get reliable statistics due to the long resolution times.

\section{Acknowledgments}
We would like to thank St\'ephane Roux for enlightening discussions,
Lo\"\i c Sorbier for his work on the $q$ algorithm and Reuven Zeitak
for a fruitful discussion on the $q$ algorithm.

\newpage
\part*{Bibliography}
\bibliographystyle{prsty} 
\bibliography{oron-herrmann}

\newpage
\part*{Tables}
\begin{table}
  $$
  \begin{array}{|rl|rl|rl|}
    \hline
    & & & & & \\
    {\dd q_1=} & {\dd -\frac{64757787245}{13047053469} \,\sqrt{3}} &
    {\dd q_2=} & {\dd - \frac{61636606607}{11780554266} \,\sqrt{3}} &
    {\dd q_3=} & {\dd - \frac{762822774179}{132529054821 } \,\sqrt{3}}
    \\ & & & & & \\
    
    {\dd q_4=} & {\dd - \frac{6707882705971}{1062824475948} \,
      \sqrt{3}} & {\dd q_5=} & {\dd
      -\frac{4445792285474}{687757544473}\,\sqrt{3}} & {\dd q_6=} &
    {\dd -\frac {2397551418033037}{365699270454792} \,\sqrt{3}}
    \\ & & & & & \\
    
    {\dd q_7=} & {\dd - \frac{892376842807618}{118456895286735}
      \,\sqrt{3}} & {\dd q_8=} & {\dd -\frac {1040529221853421}{
        136826565791760} \,\sqrt{3}} & {\dd q_9=} & {\dd- {
        \frac{10760229006102401}{1372821783991152 }} \,\sqrt{3}}
    \\ & & & & & \\
    
    {\dd q_{10}=} & {\dd - { \frac {688471809160333}{ 83961109063488}}
      \,\sqrt{3}} & {\dd q_{11}=} & {\dd - { \frac
        {8424287631067}{527998087996}} \,\sqrt{3}} & {\dd q_{12}=} &
    {\dd -{ \frac {23088074878561}{1393226957796}} \,\sqrt{3}}
    \\ & & & & & \\
    
    {\dd q_{13}=} & {\dd - { \frac {203779480235}{10761198684}}
      \,\sqrt{3}} & {\dd q_{14}=} & {\dd - { \frac
        {54690800571}{2080923200}} \,\sqrt{3}} & {\dd q_{15}=} & {\dd
      - { \frac {122907611189}{3605129400}} \,\sqrt{ 3}}
    \\ & & & & & \\
    
    {\dd q_{16}=} & {\dd - {\frac {13899925451}{389467512}}
      \,\sqrt{3}} & {\dd q_{17}=} & {\dd - {\frac
        {8553278999}{235721424}} \,\sqrt{3}} &
    {\dd q_{18}=} & {\dd - { \frac {258704085}{6896912}} \,\sqrt{3}} \\
    & & & & & \\
    
    {\dd q_{19}=} & {\dd - { \frac {15648333}{325196}} \,\sqrt{3}} &
    {\dd q_{20}=} & {\dd - { \frac {35529}{704}} \,\sqrt{3}} &
    {\dd q_{21}=} & {\dd - { \frac {199011}{3136}} \,\sqrt{3}} \\
    & & & & & \\
    
    {\dd q_{22}=} & {\dd - { \frac {2209}{33}} \,\sqrt{3}} &
    {\dd q_{23}=} & {\dd - { \frac {266}{3}} \,\sqrt{3}} & & \\
    & & & & & \\
    \hline
  \end{array}
  $$
  \caption{Values of the parameter $q$ for which the contact
    network rearranges in the {\em NT30} case for a 6 layer pile.}
  \label{tab:qlist}
\end{table}


\newpage

\newpage

\end{document}